\documentclass[prb,superscriptaddress,amsfonts,amsmath,floatfix]{revtex4}
\usepackage{graphicx}
\usepackage{bm}
\usepackage{amsmath}
\usepackage{hyperref}

\begin{document}
\title{Mathematical modeling of the gravitational field of a twisted Skyrmion string}

\author{Miftachul Hadi}
\email{itpm.id@gmail.com}
\affiliation{Department of Mathematics, Universiti Brunei Darussalam\\
Jalan Tungku Link BE1410, Gadong, Negara Brunei Darussalam}
\affiliation{Physics Research Centre, Indonesian Insitute of Sciences (LIPI)\\
Kompleks Puspiptek, Serpong, Tangerang 15314, Indonesia}
\affiliation{Department of Physics, School of Natural Sciences\\
Ulsan National Institute of Science and Technology (UNIST)\\
50, UNIST-gil, Eonyang-eup, Ulju-gun, Ulsan, South Korea}
\affiliation{Institute of Modern Physics, Chinese Academy of Sciences\\
509 Nanchang Rd., Lanzhou  730000, China}
\author{Malcolm Anderson}
\email{malcolm.anderson@ubd.edu.bn}
\affiliation{Department of Mathematics, Universiti Brunei Darussalam\\
Jalan Tungku Link BE1410, Gadong, Negara Brunei Darussalam}
\author{Andri Husein}
\email{decepticon1022@gmail.com}
\affiliation{Department of Physics, University of Sebelas Maret\\
Jalan Ir. Sutami 36 A, Surakarta 57126, Indonesia}

\date{\today}
\begin{abstract}
In this paper we study the gravitational field of a straight string generated from a class of nonlinear sigma models, specifically the Skyrme model without a twist and the Skyrme model with a twist (the twisted Skyrmion string). The twist term, $mkz$, is included to stabilize the vortex solution. To model the effects of gravity, we replace the Minkowski tensor, $\eta^{\mu\nu}$, in the standard Skyrme Lagrangian density with a space-time metric tensor, $g^{\mu\nu}$, assumed to be static and cylindrically symmetric. The Einstein equations for the metric and field components are then derived. This work is still in progress.
\end{abstract}

\maketitle

\section{INTRODUCTION TO THE $O(3)$ NONLINEAR SIGMA MODEL}
A nonlinear sigma model is an $N$-component scalar field theory in which the fields are functions defining a mapping from the space-time to a target manifold \cite{Zakrzewski}. 
By a nonlinear sigma model, we mean a field theory with the following properties \cite{hans02}:
\begin{itemize}
\item[(1)] The fields, $\phi(x)$, of the model are subject to nonlinear constraints at all points $x\in\mathcal{M}_0$, where $\mathcal{M}_0$ is the source (base) manifold, i.e. a spatial submanifold of the (2+1) or (3+1)-dimensional space-time manifold.
\item[(2)] The constraints and the Lagrangian density are invariant under the action of a global (space-independent) symmetry group, $G$, on $\phi(x)$.
\end{itemize}

The Lagrangian density of a free (without potential) nonlinear sigma model on a Minkowski background space-time is defined to be \cite{chen}
\begin{equation}\label{1}
\mathcal{L}=\frac{1}{2\lambda^2}~\gamma_{AB}(\phi)~\eta^{\mu\nu}~\partial_\mu\phi^A~\partial_\nu\phi^B
\end{equation}
where $\gamma_{AB}(\phi)$ is the field metric, $\eta^{\mu\nu}$ is the Minkowski metric tensor, $\lambda$ is a scaling constant with dimensions of (length/energy)$^{1/2}$ and $\phi={\phi^A}$ is the collection of fields. Greek indices run from 0 to $d-1$, where $d$ is the dimension of the space-time, and upper-case Latin indices run from 1 to $N$.  

The simplest example of a nonlinear sigma model is the $O(N)$ model, which consists of $N$ real scalar fields, $\phi^A$, with the Lagrangian density \cite{hans02}
\begin{equation}\label{2}
\mathcal{L}=\frac{1}{2\lambda^2}~\delta_{AB}~\eta^{\mu\nu}~\frac{\partial\phi^A}{\partial x^\mu}~\frac{\partial\phi^A}{\partial x^\nu}
\end{equation}
where the scalar fields, $\phi^A$, satisfy the constraint
\begin{equation}\label{3}
\delta_{AB}~\phi^A\phi^B=1
\end{equation}
and $\delta_{AB}$ is the Kronecker delta. The Lagrangian density (\ref{2}) is obviously invariant under the global (space independent) orthogonal transformations $O(N)$, i.e. the group of $N$-dimensional rotations \cite{hans02}
\begin{equation}\label{4}
\phi^A\rightarrow\phi'^A=O^A_B~\phi^B.
\end{equation}

One of the most interesting examples of a $O(N)$ nonlinear sigma model, due to its topological properties, is the $O(3)$ nonlinear sigma model in 1+1 dimensions, with the Lagrangian density \cite{wiki1}
\begin{equation}\label{5}
\mathcal{L}=\frac{1}{2\lambda^2}~\eta^{\mu\nu}~\partial_\mu\phi~.~\partial_\nu\phi 
\end{equation}
where $\mu$ and $\nu$ range over $\{0,1\}$, and $\phi=(\phi^1,\phi^2,\phi^3)$, subject to the constraint $\phi\cdot\phi=1$, where the dot (.) denotes the standard inner product on $R^3$. 

For a $O(3)$ nonlinear sigma model in any number $d$ of space-time dimensions the target manifold is the unit sphere $S^2$ in $R^3$, and $\mu$ and $\nu$ in the Lagrangian density (\ref{5}) run from 0 to $d-1$.
A simple representation of $\phi$ (in the general time-dependent case) is
\begin{equation}\label{6}
\phi=
\begin{pmatrix}
\sin f(t,{\bf r})~\sin g(t,{\bf r}) \\
\sin f(t,{\bf r})~\cos g(t,{\bf r}) \\
\cos f(t,{\bf r})
\end{pmatrix}
\end{equation}
where $f$ and $g$ are scalar functions on the background space-time, with Minkowski coordinates $x^\mu=(t,{\bf r})$. In what follows, the space-time dimension $d$ is taken to be 4, and so $\bf r$ is a 3-vector.

If we substitute (\ref{6}) into (\ref{5}), the Lagrangian density becomes
\begin{equation}\label{7}
\mathcal{L}=\frac{1}{2\lambda^2}[\eta^{\mu\nu}~\partial_\mu f~\partial_\nu f+(\sin^2f)~\eta^{\mu\nu}~\partial_\mu g~\partial_\nu g]
\end{equation}
The Euler-Lagrange equations associated with $\mathcal{L}$ in (\ref{7}) are
\begin{eqnarray}\label{8}
\eta^{\mu\nu}~\partial_\mu\partial_\nu f-(\sin f~\cos f)~\eta^{\mu\nu}~\partial_\mu g~\partial_\nu g=0
\end{eqnarray}
and
\begin{eqnarray}\label{9}
\eta^{\mu\nu}~\partial_\mu\partial_\nu g+2(\cot f)~\eta^{\mu\nu}~\partial_\mu f~\partial_\nu g=0.
\end{eqnarray}

\section{SOLITON SOLUTION}
Two solutions to the $O(3)$ field equations (\ref{8}) and (\ref{9}) are 
\begin{itemize}
\item[(i)] a monopole solution, which has form
\begin{equation}\label{10}
\phi=\hat{\textbf{r}}=
\begin{pmatrix}
x/\rho\\
y/\rho\\
z/\rho\\
\end{pmatrix}
\end{equation}
where $\rho=(x^2+y^2+z^2)^{1/2}$ is the spherical radius; and
\item[(ii)] a vortex solution, which is found by imposing the 2-dimensional ''hedgehog'' ansatz
\begin{equation}\label{11}
\phi=\begin{pmatrix}
\sin f(r)~\sin g(r)\\
\sin f(r)~\cos g(r)\\
\cos f(r)
\end{pmatrix}
=
\begin{pmatrix}
\sin f(r)~\sin (n\theta-\chi)\\
\sin f(r)~\cos (n\theta-\chi)\\
\cos f(r)
\end{pmatrix}
\end{equation}
where $r=(x^2+y^2)^{1/2}$, $\theta=\arctan (x/y)$, $n$ is a positive integer, and $\chi$ is a constant phase factor.
\end{itemize}
A vortex is a stable time-independent solution to a set of classical field equations that has finite energy in two spatial dimensions; it is a two-dimensional soliton. In three spatial dimensions, a vortex becomes a string, a classical solution with finite energy per unit length \cite{preskill}. Solutions with finite energy, satisfying the appropriate boundary conditions, are candidate soliton solutions \cite{manton}.

The boundary conditions that are normally imposed on the vortex solution (\ref{11}) are $f(0)=\pi$ and $\lim_{r\to\infty}f(r)=0$, so that the vortex ''unwinds'' from $\phi=-\hat{\textbf{z}}$ to $\phi=\hat{\textbf{z}}$ as $r$ increases from 0 to $\infty$. The function $f$ in this case satisfies the field equation 
\begin{equation}\label{12}
r~\frac{d^2f}{dr^2}+\frac{df}{dr}-\frac{n^2}{r}~\sin f~\cos f=0
\end{equation}
There is in fact a family of solutions to this equation satisfying the standard boundary conditions:
\begin{equation}\label{13}
\sin f=\frac{2K^{1/2}r^n}{1+Kr^{2n}}
\end{equation}
or equivalently
\begin{equation}\label{14}
\cos f=\frac{Kr^{2n}-1}{Kr^{2n}+1}
\end{equation}
where $K$ is positive constant.

The energy density of a static (time-independent) field with Lagrangian density, $\mathcal{L}$, (\ref{7}) is
\begin{eqnarray}\label{15}
-\mathcal{L}=-\frac{1}{2\lambda^2}\left[\eta^{\mu\nu}~\partial_\mu f~\partial_\nu f+(\sin^2 f)~\eta^{\mu\nu}~\partial_\mu g~\partial_\nu g\right]
\end{eqnarray}
The energy density of the monopole solution is
\begin{equation}\label{16}
-\mathcal{L}=\frac{1}{\lambda^2\rho^2}
\end{equation}
while the energy density of the vortex solutions is
\begin{eqnarray}\label{17}
-\mathcal{L}=\frac{4Kn^2}{\lambda^2}\frac{r^{2n-2}}{(Kr^{2n}+1)^2}.
\end{eqnarray}
In particular, the total energy
\begin{equation}\label{18}
E=\int\int\int (-\mathcal{L})~ dx~dy~dz,
\end{equation}
of both the monopole solution and the vortex solutions is infinite. But the energy per unit length of the vortex solutions
\begin{equation}\label{19}
\mu=\int\int (-\mathcal{L})~dx~dy=\frac{4\pi n}{\lambda^2}
\end{equation}
is finite, and does not depend on the value of $K$. (We use the same symbol for the energy per unit length and the mass per unit length, due to the equivalence of energy and mass embodied in the relation $E=mc^2$. Here, we choose units in which $c=1$). 

This last fact means that the vortex solutions in the nonlinear sigma models have no preferred scale. A small value of $K$ corresponds to a more extended vortex solution, and a larger value of $K$ corresponds to a more compact vortex solution, as can be seen by plotting $f$ (or $-\mathcal{L}$) for different values of $K$ and a fixed value of $n$. This means that the vortex solutions are what is called neutrally stable to changes in scale. As $K$ changes, the scale of the vortex changes, but the mass per unit length, $\mu$, does not. Note that because of equation (\ref{19}), there is a preferred winding number, $n=1$, corresponding to the smallest possible positive value of $\mu$.

Furthermore, it can be shown that the topological charge, $T$, of the vortex defined by
\begin{eqnarray}\label{20}
T\equiv \frac{1}{4\pi}~\varepsilon_{ABC}\int\int \phi^A~\partial_x \phi^B ~\partial_y \phi^C~dx~dy
\end{eqnarray}
where $\varepsilon_{ABC}$ is the Levi-Civita symbol, is conserved, in the sense that $\partial_t T=0$ no matter what coordinate dependence is assumed for $f$ and $g$ in (\ref{11}). So the topological charge is a constant, even when the vortex solutions are perturbed. Also, it is simply shown that for the vortex solutions 
\begin{eqnarray}\label{21}
T=-\frac{1}{2}n~[f(\infty)-f(0)]=n
\end{eqnarray}
and so the winding number is just the topological charge.

Because there is no natural size for the vortex solutions, we can attempt to stabilize them by adding a Skyrme term to the Lagrangian density \cite{malcolm}. 

\section{SKYRMION VORTEX WITHOUT A TWIST: THE SKYRME TERM}
The original sigma model Lagrangian density (with the unit sphere as target manifold) is
\begin{eqnarray}\label{22}
\mathcal{L}_1=\frac{1}{2\lambda^2}~\eta^{\mu\nu}~\partial_\mu\phi~.~\partial_\nu\phi
\end{eqnarray}
If a Skyrme term is added to (\ref{22}), the result is a modified Lagrangian density
\begin{eqnarray}\label{23}
\mathcal{L}_2
&=&\frac{1}{2\lambda^2}~\eta^{\mu\nu}~\partial_\mu\phi~.~\partial_\nu\phi -\underbrace{K_s~\eta^{\kappa\lambda}~\eta^{\mu\nu}(\partial_\kappa\phi\times\partial_\mu\phi)~.~(\partial_\lambda\phi\times\partial_\nu\phi)}_{\text{Skyrme term}} 
\end{eqnarray}
where the Skyrme term is the second term on the right hand side of (\ref{23}). 

With the choice of field representation (\ref{6}), equation (\ref{23}) becomes
\begin{eqnarray}\label{24}
\mathcal{L}_2
&=&\frac{1}{2\lambda^2}\left(\eta^{\mu\nu}~\partial_\mu f~\partial_\nu f+\sin^2f~\eta^{\mu\nu}~\partial_\mu g~\partial_\nu g  \right) \nonumber\\
&&-~K_s\left[2\sin^2f~\left(\eta^{\mu\nu}~\partial_\mu f~\partial_\nu f\right)\left(\eta^{\kappa\lambda}~\partial_\kappa g~\partial_\lambda g\right) -2\sin^2f~\left(\eta^{\mu\nu}~\partial_\mu f~\partial_\nu g\right)^2\right] 
\end{eqnarray}
The Euler-Lagrange equations generated by $\mathcal{L}_2$, namely
\begin{eqnarray}\label{25}
\partial_\alpha\left(\frac{\partial\mathcal{L}_2}{\partial(\partial_\alpha f)}\right)-\frac{\partial\mathcal{L}_2}{\partial f}=0
\end{eqnarray}
and
\begin{eqnarray}\label{26}
\partial_\alpha\left(\frac{\partial\mathcal{L}_2}{\partial(\partial_\alpha g)}\right)-\frac{\partial\mathcal{L}_2}{\partial g}=0
\end{eqnarray}
will be considered more closely in Section IV. The mass per unit length of a static vortex solution with Lagrangian density $\mathcal{L}_2$ (\ref{24}) is
\begin{eqnarray}\label{27}
\mu
&=&\int\int\left\{\frac{1}{2\lambda^2}\left[\left(\frac{df}{dr}\right)^2+\frac{n^2}{r^2}~\sin^2f\right] -2K_s~\frac{n^2}{r^2}~\sin^2f~\left(\frac{df}{dr}\right)^2\right\}r~dr~d\theta
\end{eqnarray}

Let us define a new radial coordinate
\begin{eqnarray}\label{28}
\overline{r}\equiv qr
\end{eqnarray}
where $q$ is a constant. Then
\begin{eqnarray}\label{29}
\frac{df}{dr}=\frac{\partial f}{\partial\overline{r}}~q
\end{eqnarray}
and the mass per unit length (\ref{27}) can be rewritten as
\begin{eqnarray}\label{30}
\mu
&=&\int\int\left\{\frac{1}{2\lambda^2}\left[\left(\frac{\partial f }{\partial\overline{r}}\right)^2+\frac{n^2}{\overline{r}^2}~\sin^2f\right]-2q^2~K_s~\frac{n^2}{\overline{r}^2}~\sin^2f~\left(\frac{\partial f}{\partial\overline{r}}\right)^2\right\}\overline{r}~d\overline{r}~d\theta
\end{eqnarray}

From (\ref{30}), it is clear that the energy per unit length $\mu$ goes to $-\infty$ as $q\rightarrow\infty$. This is an indication that the vortex is unstable to uniform stretching, as it would be energetically favourable for the field profile $f$ to dilate without bound and the vortex effectively to evaporate to infinity. A more rigorous statement of this property follows on from Derrick's theorem \cite{derrick}, which states that a necessary condition for vortex stability is that $(\partial \mu/\partial q)|_{q=1}=0$. It is evident that (\ref{30}) does not satisfy this criterion.

In an attempt to fix this problem, we could add a ''mass'' term $K_v(1-\hat{\textbf{z}}.\phi)$ to the Lagrangian density $\mathcal{L}_2$, where $\hat{\textbf{z}}$ is the direction of $\phi$ at $r=\infty$ (where $f=0$). The Lagrangian density
\begin{eqnarray}\label{31}
\mathcal{L}_3=\mathcal{L}_2+K_v(1-\underline{n}.\hat{\underline{\phi}})
\end{eqnarray}
corresponds to the baby Skyrmion model in equation (2.2) of \cite{piette}, p.207. The kinetic term together with the Skyrme term in $\mathcal{L}_2$ are not sufficient to stabilize a baby Skyrmion, as the kinetic term in $2+1$ dimensions is conformally invariant and the baby Skyrmion can always reduce its energy by inflating indefinitely. This is in contrast to the usual Skyrme model, in which the Skyrme term prohibits the collapse of the $3+1$ soliton \cite{gisiger}.
The mass term is added to limit the size of the baby Skyrmion. 

\section{SKYRMION VORTEX WITH A TWIST: THE TWISTED SKYRME MODEL}
Instead of adding a mass term to stabilize the vortex, we will retain the Skyrme model Lagrangian (\ref{24}) but include a twist in the field $g$ in (\ref{11}). That is, instead of choosing
\begin{equation}\label{32}
g=n\theta-\chi
\end{equation}
we choose
\begin{equation}\label{33}
g=n\theta+mkz
\end{equation}
where $mkz$ is the twist term. The Lagrangian density (\ref{24}) then becomes
\begin{eqnarray}\label{34}
\mathcal{L}_2
&=&\frac{1}{2\lambda^2}\left[\left(\frac{df}{dr}\right)^2+\sin^2f\left(\frac{n^2}{r^2}+m^2k^2\right)\right] -2K_s~\sin^2f\left(\frac{df}{dr}\right)^2\left(\frac{n^2}{r^2}+m^2k^2\right)
\end{eqnarray}
The value of the twist lies in the fact that in the far field, where $r\to\infty$ then $f\to0$, the Euler-Lagrange equations for $f$ for both $\mathcal{L}_3$ without a twist and $\mathcal{L}_2$ with a twist are formally identical to leading order, with $m^2k^2/\lambda^2$ in the twisted case playing the role of the mass coupling constant $K_v$. So it is expected that the twist term will act to stabilize the vortex just as the mass term does in $\mathcal{L}_3$.

On a physical level, the twist can be identified with a circular stress in the plane perpendicular to the vortex string (which can be imagined e.g. as a rod aligned with the $z$-axis). The direction of the twist can be clockwise or counter-clockwise. In view of the energy-mass relation, the energy embodied in the stress term contributes to the gravitational field of the string, with the net result that the trajectories of freely-moving test particles differ according to whether they are directed clockwise or counter-clockwise around the string \cite{malcolm}.

The Euler-Lagrange equation (\ref{25}) corresponding to the twisted Skyrmion vortex Lagrangian density (\ref{34}) reads
\begin{eqnarray}\label{35}
0
&=&\frac{1}{\lambda^2}\left[\frac{d^2f}{dr^2}+\frac{1}{r}~\frac{df}{dr}-\left(\frac{n^2}{r^2}+m^2k^2\right)~\sin f~\cos f\right] -4\left(\frac{n^2}{r^2}+m^2k^2\right)K_s~\sin^2f~\left(\frac{d^2f}{dr^2}-\frac{1}{r}~\frac{df}{dr}\right)\nonumber\\
&&-~4\left(\frac{n^2}{r^2}+m^2k^2\right)K_s~\sin f~\cos f~\left(\frac{df}{dr}\right)^2
\end{eqnarray}
It should be noted that the second Euler-Lagrange equation (\ref{26}) is satisfied identically if $g$ has the functional form (\ref{33}).

\section{GRAVITATIONAL FIELD OF A TWISTED SKYRMION STRING}
We are interested in constructing the space-time generated by a twisted Skyrmion string. Without gravity, the Lagrangian density of the system is $\mathcal{L}_2$, as given in equation (\ref{24}).

To add gravity, we replace $\eta^{\mu\nu}$ in $\mathcal{L}_2$ with a space-time metric tensor $g^{\mu\nu}$, which in view of the time-independence and cylindrical symmetry of the assumed vortex solution is taken to be a function of $r$ alone. Metric tensor, $g^{\mu\nu}$, is of course the inverse of the covariant metric tensor, $g_{\mu\nu}$, of the space-time where $g^{\mu\nu}=(g_{\mu\nu})^{-1}$. We use a cylindrical coordinate system $(t,r,\theta,z)$, where $t$ and $z$ have unbounded range, $r\in[0,\infty)$ and $\theta\in[0,2\pi)$. 
The components of the metric tensor
\begin{eqnarray}\label{36}
g_{\mu\nu}
=
\begin{pmatrix}
g_{tt} & 0      & 0                & 0  \\
0      & g_{rr} & 0                & 0  \\
0      & 0      & g_{\theta\theta} & g_{\theta z} \\
0      & 0     & g_{z\theta}      & g_{zz}
\end{pmatrix}
\end{eqnarray}
are all functions of $r$, and the presence of the off-diagonal components $g_{\theta z}=g_{z\theta}$ reflects the twist in the space-time.

The Lagrangian we will be using is
\begin{eqnarray}\label{37}
\mathcal{L}_4
&=&\frac{1}{2\lambda^2}~(g^{\mu\nu}~\partial_\mu f~\partial_\nu f+\sin^2f~g^{\mu\nu}~\partial_\mu g~\partial_\nu g) \nonumber\\
&&+~2K_s~\sin^2f~[(g^{\mu\nu}~\partial_\mu f~\partial_\nu f)(g^{\kappa\lambda}~\partial_\kappa g~\partial_\lambda g) -2\sin^2f~(g^{\mu\nu}~\partial_\mu f~\partial_\nu g)^2]
\end{eqnarray}
with $f=f(r)$ and $g=n\theta+mkz$.

We need to solve
\begin{itemize}
\item[(i)] the Einstein equations
\begin{eqnarray}\label{38}
G_{\mu\nu}
&=& -\frac{8\pi G}{c^4}~T_{\mu\nu}  
\end{eqnarray}
where the stress-energy tensor of the vortex, $T_{\mu\nu}$, is defined by
\begin{eqnarray}\label{39}
T_{\mu\nu}
&\equiv&2\frac{\partial\mathcal{L}_4}{\partial g^{\mu\nu}}-g_{\mu\nu}~\mathcal{L}_4
\end{eqnarray}
and
\begin{eqnarray}\label{40}
G_{\mu\nu} 
&=& R_{\mu\nu}-\frac{1}{2}g_{\mu\nu}~R 
\end{eqnarray}
with
$R^{\mu\nu}$ the Ricci tensor and 
\begin{eqnarray}\label{41}
R=g_{\mu\nu}~R^{\mu\nu} = g^{\mu\nu}~R_{\mu\nu}
\end{eqnarray}
the Ricci scalar; and
\item[(ii)] the field equations for $f$ and $g$
\begin{eqnarray}\label{42}
\nabla^\mu\frac{\partial\mathcal{L}_4}{\partial(\partial f/\partial x^\mu)}=\frac{\partial\mathcal{L}_4}{\partial f}
\end{eqnarray}
and
\begin{eqnarray}\label{43}
\nabla^\mu\frac{\partial\mathcal{L}_4}{\partial(\partial g/\partial x^\mu)}=\frac{\partial\mathcal{L}_4}{\partial g}
\end{eqnarray}
\end{itemize}
However, the field equations for $f$ and $g$ are in fact redundant, as they are satisfied identically whenever the Einstein equations are satisfied, by virtue of the Bianchi identities $\nabla_\mu G^{\mu}_{\nu}=0$. So only the Einstein equations will be considered in this section.

To simplify the Einstein equations, we first choose a gauge condition that narrows down the form of the metric tensor. The gauge condition preferred here is that
\begin{eqnarray}\label{44}
g_{\theta\theta}~g_{zz}-(g_{\theta z})^2=r^2
\end{eqnarray}
The geometric significance of this choice is that the determinant of the 2-metric tensor projected onto the surfaces of constant $t$ and $z$ is $r^2$, and so the area element on these surfaces is just $r~dr~d\theta$.

As a further simplification, we write
\begin{eqnarray}\label{45}
g_{tt}=A^2;~~~~~g_{rr}=-B^2
\end{eqnarray}
\begin{eqnarray}\label{46}
g_{\theta\theta}=-C^2;~~~~~g_{\theta z}=\omega
\end{eqnarray}
and so
\begin{eqnarray}\label{47}
g_{zz}=-\left(\frac{r^2+\omega^2}{C^2}\right).
\end{eqnarray}
The metric tensor, $g_{\mu\nu}$, therefore has the form
\begin{eqnarray}\label{48}
g_{\mu\nu}=
\begin{pmatrix}
A^2  &  0     &    0     &  0  \\
0    &  -B^2  &    0     &  0  \\
0    &  0     &  -C^2    &  \omega  \\
0    &  0     &   \omega & -\left(\frac{r^2+\omega^2}{C^2}\right)
\end{pmatrix}
\end{eqnarray}
Substituting (\ref{48}) into $\mathcal{L}_4$ gives
\begin{eqnarray}\label{49}
\mathcal{L}_4
&=&-\frac{1}{2\lambda^2}\left\{B^{-2}f'^2+\sin^2f[n^2(1+\omega^2/r^2)C^{-2} +2mkn~r^{-2}\omega+m^2k^2r^{-2}C^2]\right\}\nonumber\\
&&+~2K_s~\sin^2f\left\{B^{-2}f'^2[n^2(1+\omega^2/r^2)C^{-2} +2mkn~r^{-2}\omega+m^2k^2r^{-2}C^2]\right\} 
\end{eqnarray}
From equations (\ref{39}), (\ref{40}), (\ref{48}) and (\ref{49}), the non-zero components of the stress-energy tensor, $T_{\mu\nu}$, are:
\begin{eqnarray}\label{50}
T_{tt}
=\frac{1}{2\lambda^2}~A^2B^{-2}f'^2  -A^2\left( -\frac{1}{2\lambda^2} +2K_s B^{-2}f'^2 \right)\sin^2f[n^2(1+\omega^2r^{-2})C^{-2} +2mkn~\omega r^{-2} +(mk)^2C^2 r^{-2} ] 
\end{eqnarray}
\begin{eqnarray}\label{51}
T_{rr}
=\frac{1}{2\lambda^2}~f'^2  -B^2\left(\frac{1}{2\lambda^2}  +2K_sB^{-2}f'^2\right)\sin^2f \left[n^2(1+\omega^2r^{-2})C^{-2} 
+2mkn~\omega r^{-2} +(mk)^2C^2r^{-2}\right] 
\end{eqnarray}
\begin{eqnarray}\label{52}
T_{\theta\theta}
=-\frac{1}{2\lambda^2}~ B^{-2}C^2f'^2 +\left(-\frac{1}{2\lambda^2} +2K_sB^{-2}f'^2  \right)\sin^2f\left[ n^2(-1 +\omega^2r^{-2}) +2mkn~\omega C^2 r^{-2}  +(mk)^2r^{-2}C^4\right] 
\end{eqnarray}
\begin{eqnarray}\label{53}
T_{\theta z}
=\frac{1}{2\lambda^2}~B^{-2}\omega f'^2+\left(\frac{1}{2\lambda^2} -2K_s B^{-2}f'^2  \right)\sin^2f\left[n^2(1 +\omega^2r^{-2})\omega C^{-2}  +2mkn(1 +\omega^2r^{-2}) +(mk)^2r^{-2}\omega C^2\right]
\end{eqnarray}
and
\begin{eqnarray}\label{54}
T_{zz}
&=&-\frac{1}{2\lambda^2}~(\omega^2 +r^2)B^{-2}C^{-2}f'^2 +\left(-\frac{1}{2\lambda^2} +2K_s B^{-2}f'^2\right)\sin^2f\nonumber\\
&&\times~\left[n^2r^2 (1 +\omega^2r^{-2})^2C^{-4} +2mkn(1 +r^{-2}\omega^2)\omega C^{-2} +(mk)^2(-1 +r^{-2}\omega^2)\right] 
\end{eqnarray}
where $f'=df/dr$. Note that $T_{\theta z}$ is in general non-zero, provided that either $mk$ or $\omega$ is non-zero. In fact, $mk$ acts as a source term for $\omega$, as will be seen in equation (\ref{70}) below. The twist in the vortex is therefore solely responsible for a non-zero circular stress $T_{\theta z}$.

In component form, the Einstein equations (\ref{38}) read
\begin{eqnarray}\label{55}
G_{tt}
&=&-\frac{8\pi G}{c^4}~T_{tt};~~~~~ G_{rr}= -\frac{8\pi G}{c^4}~T_{rr}
\end{eqnarray}
\begin{eqnarray}\label{56}
G_{\theta\theta}
&=&-\frac{8\pi G}{c^4}~T_{\theta\theta};~~~~~G_{\theta z}=-\frac{8\pi G}{c^4}~T_{\theta z}
\end{eqnarray}
and
\begin{eqnarray}\label{57}
G_{zz}
&=&-\frac{8\pi G}{c^4}~T_{zz}
\end{eqnarray}

The non-zero components of the Einstein tensor, $G_{\mu\nu}$, corresponding to the metric tensor (\ref{40}) are
\begin{eqnarray}\label{58}
G_{tt}
&=&R_{tt} -\frac{1}{2}g_{tt}~R = -\frac{A^2B'}{rB^3}-\frac{A^2C'}{rB^2C}+\frac{A^2C'^2}{B^2C^2}(1+r^{-2}\omega^2) -\frac{A^2\omega\omega'C'}{r^2B^2C}+\frac{A^2\omega'^2}{4r^2B^2}
\end{eqnarray}
\begin{eqnarray}\label{59}
G_{rr}
&=&R_{rr} -\frac{1}{2}g_{rr}~R = -\frac{A'}{rA}-\frac{C'}{rC}+\frac{C'^2}{C^2}(1+r^{-2}\omega^2)-\frac{\omega\omega'C'}{r^2C} +\frac{\omega'^2}{4r^2}
\end{eqnarray}
\begin{eqnarray}\label{60}
G_{\theta\theta}
&=&R_{\theta\theta} -\frac{1}{2}g_{\theta\theta}~R \nonumber\\
&=& -\frac{C^2A'}{rAB^2}-\frac{C^2A''}{AB^2}+\frac{C^2B'}{rB^3}+\frac{C^2A'B'}{AB^3} +\frac{2CC'}{rB^2}+\frac{CC''}{B^2} -\frac{C'^2}{B^2}(2+3r^{-2}\omega^2) \nonumber\\
&&+~\frac{CA'C'}{AB^2} -\frac{CB'C'}{B^3}+\frac{3\omega\omega'CC'}{r^2B^2} -\frac{3C^2\omega'^2}{4r^2B^2}
\end{eqnarray}
\begin{eqnarray}\label{61}
G_{\theta z}
&=&R_{\theta z} -\frac{1}{2}g_{\theta z}~R \nonumber\\
&=&\frac{\omega A'}{rAB^2} -\frac{\omega'A'}{2AB^2} +\frac{\omega A''}{AB^2} -\frac{\omega B'}{rB^3} +\frac{\omega'B'}{2B^3} -\frac{\omega A'B'}{AB^3} -\frac{3\omega C'}{rB^2C} -\frac{3\omega^2\omega'C'}{r^2B^2C} \nonumber\\
&&+~\frac{3\omega C'^2}{B^2C^2}(1+r^{-2}\omega^2)+\frac{\omega'}{2rB^2}-\frac{\omega''}{2B^2} +\frac{3\omega\omega'^2}{4r^2B^2}
\end{eqnarray}
with $G_{\theta z}=G_{z\theta}$, and
\begin{eqnarray}\label{62}
G_{zz}
&=&R_{zz} -\frac{1}{2}g_{zz}~R \nonumber\\
&=&-\frac{\omega^2A'}{rAB^2C^2} -\frac{r^2A''}{AB^2C^2}(1+r^{-2}\omega^2) +\frac{r^2A'B'}{AB^3C^2}(1+r^{-2}\omega^2) -\frac{r^2A'C'}{AB^2C^3}(1+r^{-2}\omega^2) \nonumber\\
&&+~\frac{\omega^2B'}{rB^3C^2} +\frac{r^2B'C'}{B^3C^3}(1+r^{-2}\omega^2) +\frac{4\omega^2 C'}{rB^2C^3} -\frac{r^2C''}{B^2C^3}(1+r^{-2}\omega^2) -\frac{3\omega^2C'^2}{B^2C^4}(1+r^{-2}\omega^2) \nonumber\\
&&+~\frac{\omega\omega'A'}{AB^2C^2} -\frac{\omega\omega'B'}{B^3C^2} -\frac{\omega\omega'C'}{B^2C^3}(1-3r^{-2}\omega^2) -\frac{\omega\omega'}{rB^2C^2} +\frac{\omega\omega''}{B^2C^2} +\frac{\omega'^2}{4B^2C^2}(1-3r^{-2}\omega^2) 
\end{eqnarray}
Substituting (\ref{58})-(\ref{62}) and (\ref{50})-(\ref{54}) into equations (\ref{55})-(\ref{57}) now gives  
\begin{eqnarray}\label{63}
-\frac{8\pi G}{c^4}~T_{tt} &=&-\frac{A^2B'}{rB^3}-\frac{A^2C'}{rB^2C}+\frac{A^2C'^2}{B^2C^2}(1+r^{-2}\omega^2) -\frac{A^2\omega\omega'C'}{r^2B^2C}+\frac{A^2\omega'^2}{4r^2B^2} 
\end{eqnarray}
\begin{eqnarray}\label{64}
-\frac{8\pi G}{c^4}~T_{rr} &=&-\frac{A'}{rA}-\frac{C'}{rC}+\frac{C'^2}{C^2}(1+r^{-2}\omega^2)-\frac{\omega\omega'C'}{r^2C} +\frac{\omega'^2}{4r^2} 
\end{eqnarray}
\begin{eqnarray}\label{65}
-\frac{8\pi G}{c^4}~T_{\theta\theta} &=&-\frac{C^2A'}{rAB^2}-\frac{C^2A''}{AB^2}+\frac{C^2B'}{rB^3}+\frac{C^2A'B'}{AB^3} +\frac{2CC'}{rB^2}+\frac{CC''}{B^2} -\frac{C'^2}{B^2}(2+3r^{-2}\omega^2) \nonumber\\
&&+~\frac{CA'C'}{AB^2} -\frac{CB'C'}{B^3}+\frac{3\omega\omega'CC'}{r^2B^2} -\frac{3C^2\omega'^2}{4r^2B^2} 
\end{eqnarray}
\begin{eqnarray}\label{66}
-\frac{8\pi G}{c^4}~T_{\theta z}
&=&\frac{\omega A'}{rAB^2} -\frac{\omega'A'}{2AB^2} +\frac{\omega A''}{AB^2} -\frac{\omega B'}{rB^3} +\frac{\omega'B'}{2B^3} -\frac{\omega A'B'}{AB^3} -\frac{3\omega C'}{rB^2C} -\frac{3\omega^2\omega'C'}{r^2B^2C} \nonumber\\
&&+~\frac{3\omega C'^2}{B^2C^2}(1+r^{-2}\omega^2)+\frac{\omega'}{2rB^2}-\frac{\omega''}{2B^2} +\frac{3\omega\omega'^2}{4r^2B^2}  
\end{eqnarray}
and
\begin{eqnarray}\label{67}
-\frac{8\pi G}{c^4}~T_{zz} 
&=&-\frac{\omega^2A'}{rAB^2C^2} -\frac{r^2A''}{AB^2C^2}(1+r^{-2}\omega^2) +\frac{r^2A'B'}{AB^3C^2}(1+r^{-2}\omega^2) -\frac{r^2A'C'}{AB^2C^3}(1+r^{-2}\omega^2) \nonumber\\
&&+~\frac{\omega^2B'}{rB^3C^2} +\frac{r^2B'C'}{B^3C^3}(1+r^{-2}\omega^2) +\frac{4\omega^2 C'}{rB^2C^3} -\frac{r^2C''}{B^2C^3}(1+r^{-2}\omega^2) -\frac{3\omega^2C'^2}{B^2C^4}(1+r^{-2}\omega^2) \nonumber\\
&&+~\frac{\omega\omega'A'}{AB^2C^2} -\frac{\omega\omega'B'}{B^3C^2} -\frac{\omega\omega'C'}{B^2C^3}(1-3r^{-2}\omega^2) -\frac{\omega\omega'}{rB^2C^2} +\frac{\omega\omega''}{B^2C^2} +\frac{\omega'^2}{4B^2C^2}(1-3r^{-2}\omega^2) 
\end{eqnarray}

These equations can be rearranged as source equations for the second derivatives of $A$, $C$ and $\omega$ and the first derivatives of $B$ and $f$ as follows:
\begin{eqnarray}\label{68}
A''= -\frac{A'}{2r}  +\frac{A'B'}{B}  +\frac{AB'}{2rB}  +\frac{AC'}{rC}  +\frac{\omega\omega'AC'}{r^2C}  -\frac{AC'^2}{C^2}(1 +r^{-2}\omega^2 )  -\frac{A\omega'^2}{4r^2}  -\frac{\kappa}{2\lambda^2}Af'^2
\end{eqnarray}
\begin{eqnarray}\label{69}
C''
&=& -\frac{C'}{r}  +\frac{CA'}{2rA}  -\frac{A'C'}{A}  -\frac{CB'}{2rB}  +\frac{B'C'}{B}  -\frac{2\omega\omega'C'}{r^2}  
+\frac{C'^2}{C}(1 +2r^{-2}\omega^2)  +\frac{C\omega'^2}{2r^2} \nonumber\\
&&-~\kappa B^2 C^{-1}r^{-2}[(n\omega +C^2km)^2 -(nr)^2] \left(-\frac{1}{2\lambda^2}  +2K_sB^{-2}f'^2\right)\sin^2f 
\end{eqnarray}
\begin{eqnarray}\label{70}
\omega''
&=& \frac{\omega'}{r}  +\frac{\omega\omega'^2}{r^2}  +\frac{\omega A'}{rA}  -\frac{\omega'A'}{A}  -\frac{\omega B'}{rB}  +\frac{\omega'B'}{B}  -\frac{4\omega C'}{rC}  -\frac{4\omega^2\omega'C'}{r^2C}  +\frac{4\omega C'^2}{C^2}(1 +r^{-2}\omega^2)\nonumber\\
&&-2\kappa B^2 C^{-2}r^{-2}\left\{\omega[(n\omega +C^2km)^2 +(nr)^2]  +2C^2kmn~r^2\right\}\left(-\frac{1}{2\lambda^2}  +2K_sB^{-2}f'^2  \right)\sin^2f 
\end{eqnarray}
\begin{eqnarray}\label{71}
B'
&=& A^{-1}BA'  +\frac{\kappa}{\lambda^2}B^3C^{-2}r^{-1}[(n\omega +C^2km)^2  +(nr)^2]\sin^2f
\end{eqnarray}
and
\begin{eqnarray}\label{72}
f'^2
&=& \left\{\kappa^{-1}\left[\frac{A'}{rA}  +\frac{C'}{rC}  -\frac{C'^2}{C^2}(1 +r^{-2}\omega^2)  +\frac{\omega\omega'C'}{r^2C}  
-\frac{\omega'^2}{4r^2}\right]  +\frac{1}{2\lambda^2}B^2C^{-2}r^{-2}[(n\omega +C^2km)^2 +(n\omega)^2]\sin^2f\right\}\nonumber\\
&&\left\{\frac{1}{2\lambda^2}  -2K_sC^{-2}r^{-2}[(n\omega  +C^2km)^2  +(nr)^2]\sin^2f\right\}^{-1}
\end{eqnarray}
where $\kappa=8\pi G/c^4$.

In order to solve equations (\ref{68})-(\ref{72}), we require boundary conditions on the functions $A$, $B$, $C$, $\omega$ and $f$ for small $r$ and in the limit as $r$ tends to $\infty$. In particular, $\omega$ should vanish both at $r=0$ (at least as rapidly as $r^2$, so as to preserve elementary flatness on the axis of symmetry) and in the limit as $r\to\infty$ (to allow for a locally Minkowski vacuum at space-like infinity). It is evident from equation (\ref{70}) that $\omega$ will be zero everywhere unless $mk\neq0$, and so the vortex twist acts as a source for the space-time twist as mentioned above.

\section{THE EINSTEIN-HILBERT ACTION}
The Lagrangian density $\mathcal{L}_4$ of course describes the matter part of the twisted Skyrme model only. When we vary $\mathcal{L}_4$ with respect to $f$ and $g$, we obtain the Euler-Lagrange equations of motion for the twisted Skyrme model, and when we vary $\mathcal{L}_4$ with respect to $g^{\mu\nu}$ we obtain the stress-energy tensor for the twisted Skyrme model. But these are not in themselves sufficient to generate the gravitational field equations for $g_{\mu\nu}$. 

To do this we need to add to $\mathcal{L}_4$ the Hilbert Lagrangian density, which is proportional to $\sqrt{(-\Delta)}R$, where $\Delta$ is the determinant of the metric tensor, $g_{\mu\nu}$, and $R$ is the Ricci scalar. In Riemannian geometry, the Ricci scalar or scalar curvature is the simplest curvature invariant on a Riemannian manifold. It assigns to each point on a Riemannian manifold a single real number determined by the intrinsic geometry of the manifold near that point. Specifically, the scalar curvature represents the amount by which the volume of a geodesic ball in a curved Riemannian manifold deviates from that of the standard ball in Euclidean space. In two dimensions, the scalar curvature is twice the Gaussian curvature, and completely characterizes the curvature of a surface. 

In general relativity, the Ricci scalar acts as the Lagrangian density for the Hilbert action. The Euler-Lagrange equations for this Lagrangian density under variations in the metric tensor constitute the vacuum Einstein field equations. In other words, if $\mathcal{L}_4= 0$ (or if $\kappa = 0$ so that there is no coupling between gravity and matter) then the Hilbert action will give us the vacuum Einstein field equations.

The Einstein-Hilbert action
\begin{eqnarray}\label{73}
I = \int{\underbrace{\sqrt{(-g)}(R+\kappa~\mathcal{L}_4)}_{\mathcal{L}_5}~ d^{4}x}
\end{eqnarray}
(or equivalently the Lagrangian density $\mathcal{L}_5$) is a functional of two sets of fields: the Skyrmionic string fields $f$ and $g$, and the metric tensor components $g_{\mu\nu}$. But only $\mathcal{L}_4$ depends on $f$ and $g$ explicitly. This means that $\mathcal{L}_4$ depends on $f$ and $g$, but $R$ does not. Of course both $\mathcal{L}_4$ and $R$ depend on $g_{\mu\nu}$. So when we vary $\mathcal{L}_4$ with respect to $f$ and $g$, we can forget $R$ (and $\sqrt{-\Delta}$) and just vary $\mathcal{L}_4$ with respect to $f$ and $g$. This gives us the equation of motion of the string.

When we want to calculate the effect of the stress-energy content of the string on the gravitational field around it, we need to vary the Einstein-Hilbert action with respect to $g_{\mu\nu}$, or more precisely the inverse metric tensor $g^{\mu\nu}$. This means that we need to vary all the terms in $\mathcal{L}_5$ with respect to $g^{\mu\nu}$, including $R$, $\mathcal{L}_4$ and $\sqrt{-\Delta}$. This then gives us the Einstein equations, which are second-order PDEs for the components of $g_{\mu\nu}$.
\\
\begin{center}
\textbf{Acknowlegment}
\end{center}
MH thanks Professor Malcolm Anderson for his long patience and clear guidance. Thanks also to Professor Eugen Simanek, Professor Edward Witten, Professor Wojtek Zakrzewski and Professor David Tong for fruitful discussions. Thanks to Professor Yongmin Cho and Professor Pengming Zhang, who drew my attention to topological objects in two-component Bose-Einstein condensates, which act as a static version of the baby Skyrmion cosmic string, to Mr Andri Husein for numerical work and fruitful discussions, to Professor Muhaimin and Dr Irwandi Nurdin for their kind help, to the Department of Mathematics at Universiti Brunei Darussalam and the Physics Research Centre LIPI for their support and the chance to do this research, and to all my kind colleagues, especially Dr Isnaeni, Dr Danang and Fika Fardila, for their strong support in various ways. Profound gratitude to my beloved mother, Siti Ruchanah for her continuous prayer, to Ika Nurlaila for her spirit, support and inspiration, and to my beloved daughter Aliya Syauqina Hadi for her purity and naughtiness. This research was fully supported by a Graduate Research Scholarship from Universiti Brunei Darussalam (GRS UBD).


\begin{thebibliography}{9}
\bibitem{Zakrzewski}
W. J. Zakrzewski (private communication).
\bibitem{hans02}
H. J. Wospakrik, Ph.D. thesis, University of Durham, 2002.
\bibitem{chen}
C. C. Chen and T. Earnest, \textit{Nonlinear Sigma Model}, University of Illinois, 2010.
\bibitem{malcolm}
M. R. Anderson (private communication).
\bibitem{wiki1}
Nonlinear Sigma Model (Wikipedia).
\bibitem{preskill}
J. Preskill, \textit{Vortices and Monopoles} (Elsevier Science Publishers B.V., 1987)
\bibitem{manton}
N. Manton and P. Sutcliffe, \textit{Topological Solitons} (Cambridge: Cambridge Monographs on Mathematical
Physics, 2004)
\bibitem{andri}
A. S. Husein (private communication).
\bibitem{simanek}
E. Simanek, arXiv: 1001.5061v1.
\bibitem{simanek1}
E. Simanek (private communication).
\bibitem{piette}
B. M. A. G. Piette, B. J. Schroers and W. J. Zakrzewski, Nuclear Physics B {\bf 439}, 205-235 (1995).
\bibitem{gisiger}
T. Gisiger and M. B. Paranjape, Physics Letters B {\bf 384}, 207-212 (1996).
\bibitem{cho}
Y. M. Cho, H. Khim and P. M. Zhang, Physical Review A {\bf 72}, 063603 (2005). 
\bibitem{hadi}
M. Hadi, M. Anderson M and A. Husein, Journal of Physics: Conference Series \textbf{539}, 012008 (2014).
\bibitem{wiki}
Scalar Curvature (Wikipedia).
\bibitem{derrick}
G. H. Derrick, J. Mathematical Phys. {\bf 5}, 1252-1254 (1964).
\end{thebibliography}
\end{document}